# Implications of self-identified race, ethnicity, and genetic ancestry on genetic association studies in biobanks within health systems

Ruth Johnson[1,2] & Bogdan Pasaniuc[3,4,5,6,7]

[1]Department of Biomedical Informatics, Harvard Medical School, Boston, MA 02115, USA.

[2]Department of Computer Science, University of California, Los Angeles, Los Angeles, CA 90095, USA.

[3]Bioinformatics Interdepartmental Program, University of California, Los Angeles, Los Angeles, CA 90095, USA.

[4]Department of Pathology and Laboratory Medicine, David Geffen School of Medicine, University of California, Los Angeles, Los Angeles, CA 90095, USA.

[5]Department of Human Genetics, David Geffen School of Medicine, University of California, Los Angeles, Los Angeles, CA 90095, USA.

[6]Department of Computational Medicine, David Geffen School of Medicine, University of California, Los Angeles, Los Angeles, CA 90095, USA.

[7]Institute of Precision Health, University of California, Los Angeles, Los Angeles, CA 90095, USA.

## Abstract

Precision medicine aims to create biomedical solutions tailored to specific factors that affect disease risk and treatment responses within the population. The success of the genomics era and recent widespread availability of electronic health records (EHR) has ushered in a new wave of genomic biobanks connected to EHR databases (EHR-linked biobanks). This perspective aims to discuss how race, ethnicity, and genetic ancestry are currently utilized to study common disease variation through genetic association studies. Although genetic ancestry plays a significant role in shaping the genetic landscape underlying disease risk in humans, the overall risk of a disease is caused by a complex combination of environmental, sociocultural, and genetic factors. When using EHR-linked biobanks to interrogate underlying disease etiology, it is also important to be aware of how the biases associated with commonly used descent-associated concepts such as race and ethnicity can propagate to downstream analyses. We intend for this resource to support researchers who perform or analyze genetic association studies in the EHR-linked biobank setting such as those involved in consortium-wide biobanking efforts. We provide background on how race, ethnicity, and genetic ancestry play a role in current association studies, highlight considerations where there is no consensus about best practices, and provide transparency about the current shortcomings.

**Introduction**

With the simultaneous advancements of clinical information and omics-based technologies, the biomedical landscape has dramatically shifted towards a data-centric approach. Precision medicine aims to provide a more precise approach to disease diagnosis, treatment, and prevention that account for individuals' unique risk factors such as lifestyle, environment, and genetics [1–3]. The relatively recent widespread availability of electronic health records (EHR) and sharp decrease in genotyping costs have ushered in a new wave of genomic biobanks connected to medical records[6]. In this work, we define an EHR-linked biobank as a large collection of human DNA samples linked with medical records from a health system. As a first step to identify characteristics of those that confer differences in disease risk and treatment responses, patients are often separated into "descent-associated" groups. When using EHR-linked biobanks to interrogate underlying disease etiology, it is important to be aware of how the biases around commonly used descent-associated concepts, such as race and ethnicity, can propagate to downstream analyses and potentially suggest socially derived patterns as true biological signal.

This perspective discusses the implications of using self-identified race and ethnicity and genetic ancestry information in genetic association studies (such as genome- and phenome-wide association studies) for common disease in EHR-linked biobanks. While we focus on EHR-linked biobanks directly embedded within medical systems in the United States (US), many of the topics discussed are widely applicable to other biobanks. Our intended audience is anyone who performs or utilizes genetic association studies with EHR-derived phenotypes, even if race, ethnicity, or genetic ancestry are not the primary focus of the study, as these concepts play an implicit role in all studies of genetic disease risk. We approach the discussion of these topics with a forward-looking perspective and while we recognize that the field has used these concepts imprecisely in the past, we maintain that this type of information can be beneficial for scientific discovery when used appropriately. We highlight considerations where there is no consensus about best practices, provide transparency about the shortcomings of current methodologies, and provide recommendations surrounding the use of race, ethnicity, and genetic ancestry when studying common disease risk in EHR-linked biobank settings.

## The rise in EHR-linked biobanks

With the acceleration of precision medicine efforts, genomic biobanks have subsequently become a major focus in the genomic medicine field where large cohorts of individuals are sampled and hundreds or even thousands of phenotypes are measured across the entire cohort. Biobanks offer a rich datasource for hypothesis free analyses, allowing the testing of millions of genetic variants against thousands of diseases or phenotypes at scale without the need for costly recruitment studies[10]. Many of the initial genomic biobanks were population-based and primarily established in Europe, such as deCODE Genetics in Iceland (established in 1996)[11], the Estonian Biobank in Estonia (1999)[12], and the UK Biobank in the United Kingdom (2006)[13]. In these settings, individuals were sampled from a population within a specific region (often by country) and phenotype information was collected via questionnaires and national healthcare databases. These datasets are often used to study common disease variation through large-scale genome-wide association studies (GWAS) and have led to the significant acceleration of findings for common disease risk[14,15].

Unfortunately, the majority of existing studies represent individuals of European ancestries, leaving a large proportion of global genetic diversity uncharacterized[16,17]. In order for precision medicine solutions to be applied equitably, especially in genetically diverse regions such as the United States, we must catalog the role of genetic variation in disease risk across all populations[19]. Although the US currently has population-level biobank initiatives underway (e.g. the All of Us Research Program, Million Veterans Program[19,20]), a significant sampling of the US population is represented through biobanks established by healthcare systems. In particular, the geographic diversity of current healthcare biobanks in the US enables the inclusion of individuals from across the country and a wide variety of environments, socioeconomic statuses, as well as ancestral backgrounds (Figure 1).

A key benefit of healthcare system biobanks is that patients' biosamples are often linked with medical record data through electronic health records. The widespread integration of EHRs in most healthcare systems within the US (>95%) was ushered in by the Health Information Technology for Economic and Clinical Health Act of 2009 which promoted the widespread adoption of health information technology in hospitals[21]. EHRs provide real-world patient information collected during routine clinical care, and when used in tandem with patients' genetic information, can offer multidimensional insights into health and disease risk[6,22–24]. EHR-linked biobanks can be used to derive a myriad of clinical phenotypes from various EHR fields or an aggregation across multiple parts of the EHR (e.g. using a combination of medication data and clinical notes). A major advantage of EHR-linked biobanks is that all participants within a biobank are treated in the same medical system using similar procedures and standards. This is especially advantageous for studies in the US since healthcare is non-standardized and is represented as a conglomeration of various health entities. A summary of current healthcare biobanks connected with EHR databases within US medical systems is available in Figure 1 and Supplementary Table 1.

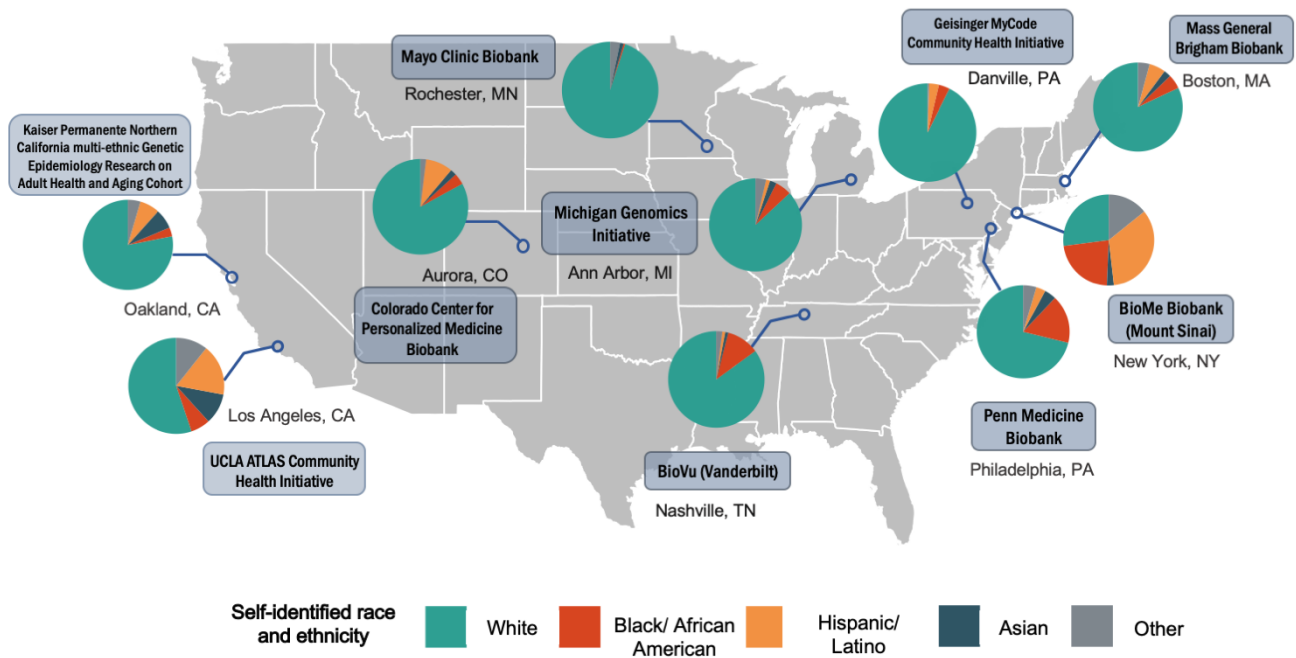

**Figure 1:** Map of EHR-linked biobanks embedded within United States healthcare systems. Pie charts show the estimated self-identified race and ethnicity percentages per biobank. We place annotations at the primary location of each health system but acknowledge that many health systems span multiple regions.

## Shortcomings and misuse of race, ethnicity, and genetic ancestry in genetic studies

Long before the rise of formal precision medicine initiatives, race and ethnicity were directly considered in many forms of clinical care management, such as the assessment of lung function, staging of chronic kidney disease, and risk estimates for atherosclerotic cardiovascular disease (ASCVD) events[36–38]. However, many of these "race-based" adjustments are sustained on the flawed assumption that race is based on biological and physiological differences, when in actuality race and ethnicity are social and power constructs created by society[39]. *Race* is defined as one's identification with a group or identity typically based on a variety of factors, including physical characteristics, social identity, and geographic history, and *ethnicity* as one's identification with a cultural group that typically shares traditions, language, and cultural norms[40](Table 1). Given the complex and fraught history of race and healthcare within the US, it is unsurprising that many current clinical guidelines treat race and ethnicity as clinical variables. For centuries, ideas surrounding "biological race" were used as tools of oppression in the institution of slavery and seizure of lands. Although these erroneous assumptions have been thoroughly disproven[41–44], they often remain an implicit belief in modern-day healthcare or are used as a shortcut for representing physiological factors in clinical medicine[45].

It has been suggested to instead use genetic ancestry to better explore how human variation affects disease risk[46]. G*enetic ancestry* is defined as the characterization of population(s) from which an individual's recent biological ancestors originated and the genetic relationship implied by the shared DNA segments inherited from these ancestors[18](Table 1). Completely eliminating the consideration of race and ethnicity in medical research could help prevent the perpetuation of the concept of biological race. However it is too simplistic to overlook the ideas of race and ethnicity that have so deeply shaped healthcare in the US[47,48], and advocates of 'race-conscious' medicine express that race and ethnicity can still provide valuable information regarding differences in structural conditions and lived experiences[49].

| Term | Definition |
| --- | --- |
| Ancestral population | a group of individuals with a common origin from which a given modern-day population has descended. These groups are characterized by shared genetics, geographical region, and time period. |
| Genetic ancestry group | a genetic characterization of a group of individuals that share a set biological ancestors and the genetic relationship implied by the shared DNA segments inherited from these ancestors |
| Genetically inferred ancestry group | an inferred genetic characterization of individuals within a group who likely share biological ancestors. This is directly dependent on the method of inference and the set of individuals in the dataset. |
| Ethnicity | a socially constructed descriptor for a cultural group characterized by, for example, shared traditions, language, and/or cultural norms |
| Race | a socially constructed descriptor based on real or perceived physical characteristics, social identity, and geographic history |

**Table 1:** Relevant glossary of terms regarding race, ethnicity, genetic ancestry, and related concepts.

This discussion of how to properly utilize race, ethnicity, and genetic ancestry information in biomedical studies has been an ongoing conversation spanning decades[40,50–55]. Recent events and initiatives have spurred renewed conversation regarding the utility of these concepts in precision medicine. First, the genetics field has come face-to-face with its Eurocentric biases and the impact of these decisions in genomics research[16,17]. The current lack of ancestral diversity in GWAS is a harbinger for healthcare disparities in these understudied populations. This concern is not unfounded given that polygenic risk scores (PRS) already exhibit poor predictive utility when models trained on individuals with primarily European genetic ancestry, which account for the vast majority of currently available PRS models[56], are applied to populations with mostly non-European genetic ancestry[19,57]. These findings have motivated large-scale government-funded projects that aim to ameliorate these disparities. In 2015, the White House announced the Precision Medicine Initiative with an initial investment of $215 million to fund research in disease prevention and treatments that take into account individual patient needs[58]. From this initiative, the All of Us Research Program was established with the goal of genotyping one million individuals across the United States with a focus on groups that have been historically underrepresented in biomedical research[59]. The National Institutes of Health (NIH) have also separately funded initiatives such as the Polygenic Risk Methods in Diverse Populations (PRIMED) Consortium which was created to develop and evaluate methods for polygenic risk prediction in populations of diverse genetic ancestry[60].

Second, there has recently been a drastic shift in the appropriation of genetics research by extremists and nationalists, twisting scientific work to claim biological justification of a racial hierarchy[61]. The exemplary sample sizes of the biobank era have also subsequently increased the number of complex traits that show a degree of genetic signal. This includes numerous studies on phenotypes relating to cognitive behavior and education[62–65], which are willingly misappropriated to support ideas regarding racial ties to IQ which have been thoroughly disproven[66,67]. However, these false ideas can eventually manifest as hatred and violence. Recent racially-motivated attacks (as of February 2023) such as those in Buffalo, NY, Christchurch, New Zealand, and El Paso, Taxes, have brought to light how modern research can be maliciously distorted[61]. The postgenomic era has granted access to copious amounts of genetics research and datasets to the public (*e.g.* publicly available data such as 1000 Genomes[68] or direct-to-consumer genetic tests). These resources have been used by racist pseudo-scientists to create new content that supports their supremacist beliefs. This content is then disseminated on online platforms where their findings claim to be supported by academic-based studies[69].

Many members of the research community have initiated conversations about potential guardrails in genetics research and scientific communication that can help prevent this misuse[70,71]. For example, although EHRs and biomedical literature often contain the term "Caucasian" as a racial category, this term has since been denounced due to its erroneous origins and historically racist implications[72], leading to a precipitous decline of the term's usage in recent biomedical publications[73]. The National Academies of Sciences, Engineering, and Medicine also recently distributed a report regarding best practices for using race, ethnicity, and other population descriptors in genomics research[7]. Another insightful review outlines how slight ambiguities in scientific writing and visualizations can lead to the misappropriation of genetic ancestry and population genomics studies[74]. Organizations within the research community, such as the American Society of Human Genetics, have also formally addressed their fraught historical ties with racism, eugenics, and systematic discrimination and are actively working towards equitable policies to prevent its resurgence[75–77]. In addition to these efforts from organizations and committees, it is also important that individual researchers understand and prevent potential misuse in their own specific studies.

## Role of genetic ancestry in genetic disease variation

Genetic ancestry has a significant impact on the genetic landscape underlying human disease risk. Differences in disease prevalence across populations are due to a complex combination of environmental, sociocultural, and genetic factors[53]. Patterns of genetic variation at a population-level are shaped through demographic, stochastic, and evolutionary events such as migration, genetic drift, and adaptive evolution[78]. Understanding these patterns within and across *ancestral populations* (a group of individuals with a common gene pool characterized by cohesive forces such as genetics and reproductive interactions) (Table 1)[79] provides an empirical basis for studying how genetic architecture influences common disease risk. We refer to genetic architecture as the characterization of the genetic contributions influencing a given phenotype, including causal variants and genetic effects, variation in allele frequencies, and linkage disequilibrium [80].

*Causal variants and effects*

Most common genetic variants (MAF > 1%) are shared across populations around the globe, an estimate which is consistent with the assumption that most common genetic variation was present before the human expansion out of Africa[53]. Overall, genetic disease risk tends to be conferred by these variants that are shared across ancestries[81]. Furthermore, the majority of significant associations from genome-wide association studies have been replicated both in magnitude and direction across multiple genetic ancestry groups [82–84]. Associations that do not replicate across ancestries could be due to differences in effects (effect size heterogeneity) at GWAS loci. Although true heterogeneity in biological effects across ancestries is possible, this has only been concluded for a small fraction of associations. Meta-analyses from the first large-scale Global Biobank Meta-Analysis Initiative study found that only 3% of genome-wide significant loci displayed significant heterogeneity in effect sizes across ancestry[85], although increased sample sizes may enable the increased statistical power needed to identify significant effect size differences. For example, a study from the PAGE Consortium consisting of ~50,000 individuals of non-European descent found that only 23% of replicated loci indicated significant effect size heterogeneity by genetic ancestry[84].

Increased sample sizes from biobanks have also enabled the power to capture small genetic effects that do not reach genome-wide significance. Prior studies have found that this extensive sharing of causal variants is not limited to GWAS risk regions, but instead is highly polygenic and extends across the entire genome[86–90]. Estimates of multi-ancestry genetic correlation ($\rho$), or the correlation of effect sizes between populations of differing genetic ancestry, consistently demonstrate $\rho>0$. This provides evidence that there is a level of shared genetic risk across genetic ancestries for most complex traits where the strength of correlation greatly varies by trait, ranging from $\rho=0.98$ for schizophrenia and $\rho=0.46$ for rheumatoid arthritis (both computed between European and East Asian genetic ancestry studies)[90]. Although effect sizes are relatively consistent across ancestries, the substantial range in correlation strength across different traits suggests some degree of effect size heterogeneity. A recent study finds that this reduction in genetic correlation is enriched in functionally important regions (e.g. UTR, histone marks), suggesting that these differences are influenced by pressures of selection[91]. Alternatively, it has been suggested that these prior estimates of effect size heterogeneity are inflated due to unmodeled differences in LD across ancestries[92].

*Minor allele frequencies*

Evolutionary forces such as mutation, recombination, and random genetic drift shape the frequency of genetic variants throughout the genome. The majority of common variants tend to be evolutionarily neutral, but these can still play a significant role in shaping the genetic landscape of common disease risk[93,94]. For example, the highest genetic differentiation between continental ancestry groups is between African and modern-day non-African populations due to the population bottleneck and resulting loss of genetic diversity associated with the initial migration out of Africa[95]. Due to the larger effective sample size of the African populations, which was less impacted by genetic drift, it is estimated that African populations have higher frequencies of ancestral risk alleles and lower frequencies of derived alleles at existing GWAS loci[96]. Therefore, if a disease study is only performed in non-African populations, the overall reduced frequency of the ancestral alleles will prohibit the detection of these associations, leaving a large proportion of global genetic variation uncharacterized. For example, when performing a GWAS with individuals from the populations of Ghana, Kenya, Nigeria, South Africa, and Uganda, novel significantly associated loci were identified despite the substantially smaller sample size of many biobanks (N<15,000). Many of these novel associations were attributable to variants that were extremely rare in non-African populations or entirely specific to African populations, underscoring the potential knowledge gaps caused by underrepresenting these populations[97].

Although a small fraction of the genome is influenced by evolutionary adaptive forces[93,94], recent work has suggested that the genetic landscape for many common diseases has at least been modestly affected by negative selection[98–101]. For example, a prior study found that 155 complex traits (including disease, anthropometric, and cognition traits) from the UK Biobank exhibited significant signatures of negative selection[98]. Multiple studies have also found that these acts of negative selection are enriched in biologically important regions of the genome such as in enhancer and coding regions [98,100–102]. One example of this phenomenon is the *MC1R* gene which regulates skin pigmentation and confers an increased risk of melanoma. Although African populations are typically more genetically polymorphic than non-African populations, it has been widely observed that there is reduced *MC1R* sequence diversity for individuals with African ancestry compared to other ancestries[103,104]. It is hypothesized that this is a result of negative selection where genetic variation associated with skin cancer was reduced due to the functional constraints of high ultraviolet radiation environments in parts of Africa[85,103,105,106]. This is just one example demonstrating how selective forces can uniquely shape common variation in regions of the genome.

*Linkage disequilibrium*

Linkage disequilibrium (LD) is the non-random association of alleles across the genome due to recombination[107]. Recombination occurs when contiguous segments of the genome are broken up and recombined, creating a new mosaic of haplotype blocks. These crossover events mostly occur at recombination "hotspots" which tend to occur at similar locations within a population but are typically different across populations[108]. This process creates correlations between variants within each block, but this correlation tends to decay as a function of generations due the persistent partitioning due to repeated recombination events. These recombination events inherently encode unique patterns of genetic variation within a population where the strength of correlation between variants, length of LD blocks, and location within the genome often vary across populations. For example, LD blocks within African populations tend to be shorter and have decreased levels of correlation due to the continuous admixture of ancestral populations that remained in Africa[109–111].

Considerations regarding LD are particularly important when utilizing genetic data captured on genotype arrays. To reduce the cost of genotyping, arrays only measure a subset of variants within the genome. Variants that adequately tag other nearby variants linked by LD are specifically captured on arrays and the unmeasured variants are then imputed using haplotype-based imputation panels. However, the correlations induced by LD can affect the genetic effect sizes estimated from GWAS. When comparing GWAS associations from different ancestries, the observed effect sizes may appear to have varying magnitudes (or direction) between groups due to population differences in LD. However, this type of observed effect size heterogeneity is a statistical artifact that arises from using a marginal association test on correlated variants as opposed to a true biological difference of genetic effects between populations. This is especially relevant when there are ancestry-specific differences in LD between the unobserved causal variant(s) and the observed associated variant(s) inferred from GWAS[107,112,113]. Another complication regarding LD is that GWAS only identifies associated genomic regions as opposed to individual-variant level associations since all variants within a locus are correlated. Fine-mapping aims to statistically untangle correlations due to LD in an effort to identify a set of candidate causal variants[107,112]. Because LD tends to be different across ancestries, fine-mapping can leverage the observed differences in correlations to better distinguish top variants (assuming the causal variants are shared between ancestries) where trans-ancestry fine-mapping efforts have further refined GWAS associations to sets of prospective causal variants[90,114,115].

**Genetic ancestry is distinct from self-identified race and ethnicity**

For EHR-linked biobanks with DNA samples, genetic ancestry is a favored method for describing human variation due to the strong correlation between patterns of common genetic variation and genetic ancestry. Since true ancestral information is often not available in EHR-linked biobanks, *genetically inferred ancestry* (GIA), or the inferred genetic characterization of individuals within a group who likely share recent biological ancestors, is often used instead(Table 1). GIA is dependent on the specific inference processes performed (*e.g.* PCA and clustering) and the choice of reference data used to compare samples. A predominantly used technique to characterize genetic ancestry is through principal component analysis (PCA)[116]. Specifically, PCA is performed on the matrix of individuals' genotypes where the largest principal components tend to capture variation due to ancestry-specific differences in allele frequencies[117]. Projecting the data into 2D space shows the continuous spatial variation among samples which is highly correlated with global geography[118]. Clustering algorithms (e.g. K-nearest neighbors) can be applied post-hoc to define groups of individuals based on similar patterns of spatial variation. This procedure largely eliminates the dependence on self-reported participant information and is commonly used to model genetic population structure in association studies.

GIA is often fitted against ancestral populations characterized by continental boundaries such as the "superpopulations" present in 1000 Genomes (e.g. European, African)[68]. However, ideas of race are often conflated with a continental-level representation of genetic ancestry since society tends to also delineate race by continental borders. Additionally, because race and ethnicity information is already available in the EHR, studies have used this information as a proxy for genetic ancestry. However, this substitution further perpetuates the misinterpretation of social concepts as true biology. A survey of previously published studies involving human health revealed that 49% of studies used the term "ancestry" when referring to non-genetic data and that "continental ancestry" was the most commonly utilized grouping criteria[119]. This construct conflation can lead to the inadequate profiling of disease risk and flawed conclusions regarding differences in biological signals. It has therefore been proposed to shift to descriptors of *genetic similarity* as opposed to genetic ancestry groups when describing patterns derived from these types of statistical analyses[7,120].

PCA plots used to model genetic ancestry should be interpreted with caution. "Synthetic maps" created by projecting geography onto PCs can lead to false conclusions regarding the determination of a given individual's biological origins. These PCA and clustering methods do not provide an absolute assignment of an individual's true biological origins, but rather can be viewed as a statistical methodology used to estimate patterns of similarity between genotypes. A more pernicious assumption is that an individual's race can be inferred from PC patterns. These erroneous assumptions have resulted in dangerous claims, such as the determination of Jewish heritage based on genetic information alone[121] even though the idea of a distinct genetic signature of Jewish descent has been widely refuted by numerous studies[122–124]. We emphasize that extreme caution and consideration be taken when communicating these observations so as to not insinuate connections to biological determinism, especially given the willful appropriation of scientific literature by the White nationalist community[125].

Despite being separate concepts, genetic ancestry and race and ethnicity still maintain a level of correlation due to the shared demographic and historical events that have shaped human populations, societies, and cultures. Prior studies have shown that the evolution of language broadly follows human evolution and that some demographic factors influencing genetic variation (*e.g.* mating, migration) also shape the ethnolinguistic diversity of a population[126–128]. Variation in skin pigmentation is largely explained by genetic factors, where patterns in the genome were largely shaped by adaptive selection due to differences in UV exposure[129–131]. Furthermore, sociocultural factors such as endogamy associated with religion and other cultural practices lead to assortative mating which limits gene flow between different populations. Extensive endogamy can lead to an increased disease burden due to the effects of genetic drift within a reduced effective population size. For example, within the Amish community (particularly the Old Order Amish), autosomal recessive disorders, such as spinal muscular atrophy and various forms of dwarfism, occur at an increased incidence due to the high frequency of certain founder mutations[132,133]. This has led to the development of extensive carrier screening panels for these mutations and the establishment of several clinics that specifically focus on the research, treatment, and outreach regarding these disorders within the Amish community, such as the Plain Community Translational Medicine Program and Clinic at UPMC Children's Hospital of Pittsburgh[134] and the nonprofit Clinic for Special Children[132,135].

Despite the observed correlations between race, ethnicity, and genetic ancestry, self-reported demographic information is not a direct substitution or valid proxy for describing human genetic variation. For example, information on individuals' country of origin or parents'/grandparents' country of origin have been used as an alternative for representing genetic ancestry. This assumes that an individual's reported country of origin adequately captures the genetic variation present in a country. However, the genetic profile of all individuals within a country is not homogeneous, and for regions with a history of cross-continental immigration and exogamy, immediate family origin information might not wholly reflect the genetic ancestry of individuals[33,136]. Physical characteristics, such as skin pigmentation, often play a role in an individual's racial identity, but numerous studies have shown that skin pigmentation is a poor estimator of genetic ancestry itself [137–139]. Additional studies have also shown that an individual's chosen language is not a consistent identifier in the EHR and may be biased due to concerns around stigma[140]. Thus, we emphasize that even if natural phenomena, such as genetic drift and selection, influence traits that are connected with current ideas surrounding race and ethnicity (*e.g.* skin pigmentation, religion), this does not connote a biological connection between these ideas and genetic ancestry. We encourage the use of genetic ancestry when describing biological concepts, such as genetic disease risk, and the use of race and ethnicity when discussing non-genetic, sociocultural factors.

**Performing genetic association tests**

EHR-linked biobanks present unique opportunities to conduct association studies across thousands of clinical phenotypes at scale across multiple genetic ancestry groups. A major concern for both genome- and phenome-wide association studies is properly controlling for population stratification. Biobanks should also take caution when performing MAF-based quality control procedures to prevent the filtering of variants that show highly variable MAF across ancestral populations. Quantity control metrics that are sensitive to variable allele frequencies, such as estimates of individual heterozygosity and tests of Hardy-Weinberg equilibrium, can be performed after first grouping individuals by genetically inferred ancestry. The choice of imputation reference panel can also greatly affect the accuracy of imputed variants. This is especially relevant for samples of African or Hispanic/Latino ancestry due to the unique LD patterns associated with these populations. More recent efforts, such as the TOPMed Project[141], have provided large, ancestrally diverse, whole genome sequencing panels that substantially improve imputation quality in under-represented admixed populations compared to prior panels that predominantly used samples of European ancestry[142]. A thorough discussion on recommended workflows and methodological considerations for performing GWAS in ancestrally diverse populations is provided in Peterson et al. [143].

To prevent spurious associations due to genome-wide variation associated with differences in genetic ancestry, as opposed to disease status, individuals are often grouped by GIA, where continental-level GIA is one of the most commonly used criteria. Oftentimes GIA is assigned through PCA and subsequent clustering procedures where genomic reference panel samples are used to describe each cluster. This is a preferred procedure for characterizing GIA since it eliminates the reliance on patients' self-reported information. The precision to which self-reported information effectively models population structure information likely varies by cohort, but prior biobanking studies have found evidence that population structure cannot reliably be accounted for by questionnaire data alone[136]. Furthermore, because race and ethnicity are not standardized across all healthcare systems and often vary across geographic regions, both within the US and internationally, combining results across multiple biobanks can lead to false positives due to unaccounted artifacts of population structure between biobanks[144]. Analyses would also inherently be limited to the predetermined list of self-identifying fields available in the EHR and their use can potentially reinforce biological ties with social constructs.

A key disadvantage of this grouped approach is finding an appropriate protocol to group together individuals (e.g. choice of reference panel) as well as the exclusion of individuals that could not be assigned to a specific GIA. Instead, linear mixed models (LMM) avoid this by jointly modeling all genotypes to account for population structure and cryptic relatedness, bypassing the need for ad-hoc categories[146–148]. However, jointly modeling all individuals causes great computational concerns for large biobanks. Recent methods such as BOLT-LMM[149,150] have enabled scalable association statistic estimation for quantitative traits by modeling SNP effects within a Bayesian framework. Additionally, the SAIGE software employs a saddle-point approximation to enable scalable inference for case-control phenotypes[151]. Another method called SUGEN[152] utilizes generalized estimating equations to account for non-random sampling designs and intricate relatedness between participants, both of which are major concerns when modeling biobank data. Prior studies have shown considerable success in collectively modeling all individuals in large cohorts without an increase in Type I error regardless of genetic ancestry[84,153], although there is still active concern regarding the use of LMMs as the main procedure for accounting for population structure[154–157]. A more detailed primer on methods development for EHR-linked biobanks can be found in Wolford et al.[158]

**Racial bias in EHR-derived phenotyping**

The effects of racial bias can propagate to association studies when constructing phenotypes and study cohorts from the EHR. Diseases that are consistently under- or overdiagnosed in racial or ethnic groups can translate to a loss of power in GWAS since a considerable number of individuals with a disease will be mistakenly labeled as controls[159]. For example, although Black patients are diagnosed with schizophrenia four times more frequently than White patients[160], one of the largest schizophrenia studies of participants with African ancestry from the Million Veterans Program revealed that the prevalence of schizophrenia for participants was not associated with individuals' PRS risk strata[161]. This discrepancy suggests that common genetic variation may not be the primary driver of the high schizophrenia prevalence in Black populations. Subsequent hypotheses suggest that the increased diagnostic rate could be due to clinician bias or a misdiagnosis of mood disorders[162,163]. This is just one example highlighting the numerous entry points for algorithmic bias and its potential prolific downstream effects of resulting conclusions regarding disease etiology.

Clinical notes and natural language processing (NLP) are among the most common data sources used when constructing phenotypes from the EHR[164], including those outlined in the Phenotype KnowledgeBase, a collection of validated EHR phenotyping algorithms[164].

However, racial biases inherently ingrained in clinical notes can also lead to imprecise phenotyping, driven by circumstance of bias instead of true disease presentation. For example, prior studies show that stigmatizing language appears more frequently in admission notes for non-Hispanic Black patients[165]. Another study illustrated that an NLP model for disease-related automated question answering demonstrated significant racial bias for pain mitigation, in which clinical scenarios with Black patients were more likely to be refused pain treatment compared to White patients[166]. Racial disparities in pain treatment stems from the completely speculative yet widely accepted belief that Black individuals have higher thresholds for pain due to their skin thickness[167]. This emphasizes the need for formal, transparent assessments of bias and fairness on NLP models to screen for patterns and predictions that reinforce health inequities[168–170].

However, it is typically not known a priori whether an algorithmically constructed phenotype is biased nor the exact source of the bias. As a first step, basic summary statistics of the case-control cohorts can potentially reveal hidden biases within the data if there is a strong association between non-biological factors and case-control status. Significant differences in income level, educational attainment, and geographic location between cases and controls could be indicative of non-biological confounding factors. A recent work by Dueñas et al. recommends reducing sources of heterogeneity in the phenotyping population to mitigate bias [171]. For example, due to the known biases in schizophrenia diagnoses, phenotypes could be separately constructed in groups of White, Black, and Hispanic/Latino individuals. Additionally, assessing phenome-wide differences across cases can help identify systematic differences in case-control assignments[171]. For many diseases it is expected that comorbidity profiles among cases should be relatively similar (e.g. medications, symptoms). Thus, large discrepancies between the comorbidities of cases could be indicative of larger, systematic differences in phenotyping.

**Race and ethnicity are imprecise proxies for environmental factors**

Because most genetic association testing frameworks assume that the environmental component (non-genetic factor) of a phenotype is shared across individuals, there is considerable concern that unmodeled differences in environment can induce spurious associations. A common strategy to account for this population stratification is to include self-identified race and ethnicity (SIRE) as a covariate in the model. However, this relies on the key assumption that SIRE is a suitable surrogate for systematic differences in environment. Additionally, SIRE information derived directly from the EHR provides very limited granularity of information. Race and ethnicity are typically recorded in the EHR using structured data fields where selections are limited to a list of controlled vocabularies[172] for which most EHR systems do not allow the selection of multiple race or ethnicity values or the use of freeform text to describe an individual's identity. Historically, individuals of mixed-race are assigned minority status despite their actual percentage of genetic ancestry[173,174]. This practice of hypodescent derives from the long-lasting effects of the "one-drop" rule in the United States where individuals with any perceived African lineage were subjected to racial segregation[175]. Furthermore, it cannot be wholly assumed that information reported in the EHR was provided directly by the patient, rather clinical staff and healthcare providers may directly enter patients' race and ethnicity information without direct input from the patient[176]. There is also apprehension that using SIRE as an adjustment factor can misleadingly assign race and ethnicity as a contributing cause of disease. To prevent the insidious integration of socially constructed biases into healthcare, it is crucial to acknowledge and account for the biases and complexities associated with the secondary use of EHR for research.

Prior EHR studies have offered innovative workflows to extract epidemiological variables or surrogate variables from the EHR that can be used to model more specific environmental factors[181,182]. Zip codes can be extracted from the EHR and then augmented with external datasets to construct a myriad of more specific proxy measurements for specific variables of interest. For example, measurements of socioeconomic data per zip code such as median household income, percent unemployment, and percent below poverty line, can be extracted from external demographic databases[183]. Resident addresses extracted from EHRs in tandem with geographic information system (GIS) technology can also provide precise measurements of human exposome, including measurements of pollution and environmental exposures (*e.g.* chemicals)[184]. The Institute of Medicine conducted the *Capturing Social and Behavioral Domains in Electronic Health Records* study in an effort to improve how social determinants of health (SDoH) are represented in the EHR. Specifically, the study recommended that a set of social and behavioral domains such as educational attainment, stress, physical activity, and neighborhood be integrated as fields within the EHR[185,186]. Continuing efforts to include measurements of SDoH directly into the EHR are already underway and have already shown great promise in increasing EHR-based clinical risk model performance[181,187–189].

**Recommendations**

The fast-approaching reality of personalized medicine exposes the need to reevaluate how we utilize race, ethnicity, and genetic ancestry in current-day research studies. As genomics-based diagnoses and treatments are increasingly integrated into the healthcare community, it is important to be aware of how these concepts have been used in the past and consistently reevaluate the role and impact of these ideas as research continues to evolve. We provide a list of five recommendations, though not exhaustive, regarding the use of race, ethnicity, and genetic ancestry when performing genetic association studies in EHR-linked biobank studies:

1. *Awareness of the history of race and ethnicity in healthcare.*

   Understanding the history of race and ethnicity in healthcare is critical to understanding how these concepts are used in the present-day. Many of the effects of racial biases in healthcare are implicitly ingrained into the data which can lead to the misinterpretation of associations as true biological signals. It is important that researchers know the origins of these implicit biases so they can recognize and actively address these patterns within the data.

2. *Genetic ancestry inference and other measurements of population structure should utilize genetic information instead of self-identified information.*

   Although EHR contains self-identified demographic information (e.g. race, ethnicity), this information should not be solely used to describe human genetic variation as it reinforces the false connection between social constructs or cultural factors as being biologically meaningful. Additionally, self-identified questionnaire information is not a robust proxy for genetic variation and can lead to genetic false positives due to unmodeled population structure. A recent report from the Committee on the Use of Race, Ethnicity, and Ancestry as Population Descriptors in Genomics Research provides an in-depth discussion on this topic in the broader context of the genetics field[7].

3. *Avoid using race and ethnicity as a proxy for environmental effects and instead include more precise measurements of the factors of interest.*

The use of race and ethnicity as a covariate representing environmental effects relies on the assumption that all individuals within a given group experience the same environmental factors. Instead, the EHR provides other data fields that could be utilized to capture the precise factors of interest. If it is not possible to use alternative measurements, it is essential to clearly state what information race and ethnicity is expected to capture and how those factors explicitly relate to the phenotype of interest.

4. *Test for potential racial biases in cohort ascertainment and phenotype construction.*

Not only does cohort construction and phenotyping greatly affect the statistical power of a study, these factors also have a direct impact on who is represented in the study. Because this information is derived from the EHR, racial biases in healthcare can systematically lead to the exclusion of groups of individuals even if race and ethnicity are not explicitly considered when creating cohorts and assigning phenotypes. It is important to formally screen data derived from the EHR for patterns that reinforce health inequities and report any potential concerns of bias in the study.

5. *Transparency, completeness, and rationale.*

To prevent the misinterpretations of findings, it is paramount that analyses are transparent about how information on race, ethnicity, and genetic ancestry are used in a study. It is equally important to explicitly present the rationale behind these choices as well as thoroughly document any potential drawbacks. There will never be a "perfect" way to utilize these concepts in a study, but it is more important to be clear about how each step was performed as the lack of detail can often lead to confusion and subsequent false assumptions regarding the findings.

**References**

1. Denny JC, Collins FS. Precision medicine in 2030-seven ways to transform healthcare. Cell. 2021;184: 1415–1419.

2. Lu CY, Terry V, Thomas DM. Precision medicine: affording the successes of science. NPJ Precis Oncol. 2023;7: 3.

3. Akhoon N. Precision Medicine: A New Paradigm in Therapeutics. Int J Prev Med. 2021;12: 12.

4. Aronson SJ, Rehm HL. Building the foundation for genomics in precision medicine. Nature. 2015;526: 336–342.

5. Khoury MJ, Armstrong GL, Bunnell RE, Cyril J, Iademarco MF. The intersection of genomics and big data with public health: Opportunities for precision public health. PLoS Med. 2020;17: e1003373.

6. Abul-Husn NS, Kenny EE. Personalized Medicine and the Power of Electronic Health Records. Cell. 2019;177: 58–69.

7. National Academies of Sciences, Engineering, Medicine. Using Population Descriptors in Genetics and Genomics Research: A New Framework for an Evolving Field. 2023.

8. Hollinger DA. National Culture and Communities of Descent. Rev Am Hist. 1998;26: 312–328.

9. Chandra K. WHAT IS ETHNIC IDENTITY AND DOES IT MATTER? Annu Rev Polit Sci. 2006;9: 397–424.

10. Beesley LJ, Salvatore M, Fritsche LG, Pandit A, Rao A, Brummett C, et al. The emerging landscape of health research based on biobanks linked to electronic health records: Existing resources, statistical challenges, and potential opportunities. Stat Med. 2020;39: 773–800.

11. Gulcher JR, Stefánsson K. The Icelandic Healthcare Database and informed consent. N Engl J Med. 2000;342: 1827–1830.

12. Leitsalu L, Haller T, Esko T, Tammesoo M-L, Alavere H, Snieder H, et al. Cohort Profile: Estonian Biobank of the Estonian Genome Center, University of Tartu. Int J Epidemiol. 2015;44: 1137–1147.

13. Bycroft C, Freeman C, Petkova D, Band G, Elliott LT, Sharp K, et al. The UK Biobank resource with deep phenotyping and genomic data. Nature. 2018;562: 203–209.

14. Glynn P, Greenland P. Contributions of the UK biobank high impact papers in the era of precision medicine. Eur J Epidemiol. 2020;35: 5–10.

15. Coppola L, Cianflone A, Grimaldi AM, Incoronato M, Bevilacqua P, Messina F, et al. Biobanking in health care: evolution and future directions. J Transl Med. 2019;17: 172.

16. Popejoy AB, Fullerton SM. Genomics is failing on diversity. Nature. 2016. pp. 161–164.

17. Sirugo G, Williams SM, Tishkoff SA. The Missing Diversity in Human Genetic Studies. Cell. 2019;177: 26–31.

18. Mathieson I, Scally A. What is ancestry? PLoS Genet. 2020;16: e1008624.

19. Martin AR, Kanai M, Kamatani Y, Okada Y, Neale BM, Daly MJ. Clinical use of current polygenic risk scores may exacerbate health disparities. Nat Genet. 2019;51: 584–591.

20. Gaziano JM, Concato J, Brophy M, Fiore L, Pyarajan S, Breeling J, et al. Million Veteran Program: A mega-biobank to study genetic influences on health and disease. J Clin Epidemiol. 2016;70: 214–223.

21. Adoption of Electronic Health Record Systems among U.S. Non-Federal Acute Care Hospitals: 2008-2015. [cited 3 Mar 2023]. Available: https://www.healthit.gov/data/data-briefs/adoption-electronic-health-record-systems-among-us-non-federal-acute-care-1

22. Denny JC. Chapter 13: Mining Electronic Health Records in the Genomics Era. PLoS Comput Biol. 2012;8: e1002823.

23. Zhou W, Kanai M, Wu K-HH, Rasheed H, Tsuo K, Hirbo JB, et al. Global Biobank Meta-analysis Initiative: Powering genetic discovery across human disease. Cell Genomics. 2022;2. doi:10.1016/j.xgen.2022.100192

24. COVID-19 Host Genetics Initiative. Mapping the human genetic architecture of COVID-19. Nature. 2021. doi:10.1038/s41586-021-03767-x

25. Belbin GM, Cullina S, Wenric S, Soper ER, Glicksberg BS, Torre D, et al. Toward a fine-scale population health monitoring system. Cell. 2021;184: 2068–2083.e11.

26. Pulley J, Clayton E, Bernard GR, Roden DM, Masys DR. Principles of human subjects protections applied in an opt-out, de-identified biobank. Clin Transl Sci. 2010;3: 42–48.

27. Wiley LK, Shortt JA, Roberts ER, Lowery J, Kudron E, Lin M, et al. Building a vertically-integrated genomic learning health system: The Colorado Center for Personalized Medicine Biobank. bioRxiv. 2022. doi:10.1101/2022.06.09.22276222

28. Carey DJ, Fetterolf SN, Davis FD, Faucett WA, Kirchner HL, Mirshahi U, et al. The Geisinger MyCode community health initiative: an electronic health record-linked biobank for precision medicine research. Genet Med. 2016;18: 906–913.

29. Manickam K, Buchanan AH, Schwartz MLB, Hallquist MLG, Williams JL, Rahm AK, et al. Exome Sequencing–Based Screening for BRCA1/2 Expected Pathogenic Variants Among Adult Biobank Participants. JAMA Netw Open. 2018;1: e182140–e182140.

30. Olson JE, Ryu E, Johnson KJ, Koenig BA, Maschke KJ, Morrisette JA, et al. The Mayo Clinic Biobank: a building block for individualized medicine. Mayo Clin Proc. 2013;88: 952–962.

31. Boutin NT, Schecter SB, Perez EF, Tchamitchian NS, Cerretani XR, Gainer VS, et al. The Evolution of a Large Biobank at Mass General Brigham. J Pers Med. 2022;12. doi:10.3390/jpm12081323

32. Zawistowski M, Fritsche LG, Pandit A, Vanderwerff B, Patil S, Schmidt EM, et al. The Michigan Genomics Initiative: A biobank linking genotypes and electronic clinical records in Michigan Medicine patients. Cell Genom. 2023;3: 100257.

33. Banda Y, Kvale MN, Hoffmann TJ, Hesselson SE, Ranatunga D, Tang H, et al. Characterizing Race/Ethnicity and Genetic Ancestry for 100,000 Subjects in the Genetic Epidemiology Research on Adult Health and Aging (GERA) Cohort. Genetics. 2015;200: 1285–1295.

34. The UCLA ATLAS Community Health Initiative: Promoting precision health research in a diverse biobank. Cell Genomics. 2023;3: 100243.

35. Verma A, Damrauer SM, Naseer N, Weaver J, Kripke CM, Guare L, et al. The Penn Medicine BioBank: Towards a Genomics-Enabled Learning Healthcare System to Accelerate Precision Medicine in a Diverse Population. J Pers Med. 2022;12. doi:10.3390/jpm12121974

36. Braun L. Race, ethnicity and lung function: A brief history. Can J Respir Ther. 2015;51: 99–101.

37. Diao JA, Inker LA, Levey AS, Tighiouart H, Powe NR, Manrai AK. In Search of a Better Equation - Performance and Equity in Estimates of Kidney Function. N Engl J Med. 2021;384: 396–399.

38. Subcommittee on Urinary Tract Infection, Steering Committee on Quality Improvement and Management, Roberts KB. Urinary tract infection: clinical practice guideline for the diagnosis and management of the initial UTI in febrile infants and children 2 to 24 months. Pediatrics. 2011;128: 595–610.

39. Kendi IX. How to Be an Antiracist. One World; 2019.

40. Bamshad M, Wooding S, Salisbury BA, Stephens JC. Deconstructing the relationship between genetics and race. Nat Rev Genet. 2004;5: 598–609.

41. Koenig BA, Lee SS-J, Richardson SS, Richardson S. Revisiting Race in a Genomic Age. Rutgers University Press; 2008.

42. Sussman RW. The Myth of Race: The Troubling Persistence of an Unscientific Idea. 2016.

43. Fuentes A, Ackermann RR, Athreya S, Bolnick D, Lasisi T, Lee S-H, et al. AAPA Statement on Race and Racism. Am J Phys Anthropol. 2019;169: 400–402.

44. Umek W, Fischer B. We Should Abandon "Race" as a Biological Category in Biomedical Research. Female Pelvic Med Reconstr Surg. 2020;26: 719–720.

45. Cerdeña JP, Asabor EN, Plaisime MV, Hardeman RR. Race-based medicine in the point-of-care clinical resource UpToDate: A systematic content analysis. EClinicalMedicine. 2022;52: 101581.

46. Borrell LN, Elhawary JR, Fuentes-Afflick E, Witonsky J, Bhakta N, Wu AHB, et al. Race and Genetic Ancestry in Medicine — A Time for Reckoning with Racism. New England Journal of Medicine. 2021. pp. 474–480. doi:10.1056/nejmms2029562

47. Yearby R, Clark B, Figueroa JF. Structural Racism In Historical And Modern US Health Care Policy. Health Aff . 2022;41: 187–194.

48. CDC. Racism and Health. In: Centers for Disease Control and Prevention [Internet]. 24 May 2022 [cited 6 Mar 2023]. Available: https://www.cdc.gov/minorityhealth/racism-disparities/index.html

49. From race-based to race-conscious medicine: how anti-racist uprisings call us to act. Lancet. 2020;396: 1125–1128.

50. Foster MW, Sharp RR. Race, ethnicity, and genomics: social classifications as proxies of biological heterogeneity. Genome Res. 2002;12: 844–850.

51. Ioannidis JPA, Powe NR, Yancy C. Recalibrating the Use of Race in Medical Research. JAMA. 2021;325: 623–624.

52. Hunt LM, Megyesi MS. The ambiguous meanings of the racial/ethnic categories routinely used in human genetics research. Soc Sci Med. 2008;66: 349–361.

53. Kittles RA, Weiss KM. Race, ancestry, and genes: implications for defining disease risk. Annu Rev Genomics Hum Genet. 2003;4: 33–67.

54. Sirugo G, Tishkoff SA, Williams SM. The quagmire of race, genetic ancestry, and health disparities. J Clin Invest. 2021;131. doi:10.1172/JCI150255

55. Website. Available: https://www.nejm.org/doi/full/10.1056/NEJMms2029562

56. Lambert SA, Gil L, Jupp S, Ritchie SC, Xu Y, Buniello A, et al. The Polygenic Score Catalog as an open database for reproducibility and systematic evaluation. Nat Genet. 2021;53: 420–425.

57. Martin AR, Gignoux CR, Walters RK, Wojcik GL, Neale BM, Gravel S, et al. Human Demographic History Impacts Genetic Risk Prediction across Diverse Populations. Am J Hum Genet. 2017;100: 635–649.

58. White House Precision Medicine Initiative. In: The White House [Internet]. [cited 3 Mar 2023]. Available: https://obamawhitehouse.archives.gov/node/333101

59. All of Us Research Program Investigators, Denny JC, Rutter JL, Goldstein DB, Philippakis A, Smoller JW, et al. The "All of Us" Research Program. N Engl J Med. 2019;381: 668–676.

60. PRIMED consortium. [cited 5 Apr 2023]. Available: https://primedconsortium.org/

61. Panofsky A, Dasgupta K, Iturriaga N. How White nationalists mobilize genetics: From genetic ancestry and human biodiversity to counterscience and metapolitics. Am J Phys Anthropol. 2021;175: 387–398.

62. Lam M, Trampush JW, Yu J, Knowles E, Davies G, Liewald DC, et al. Large-Scale Cognitive GWAS Meta-Analysis Reveals Tissue-Specific Neural Expression and Potential Nootropic Drug Targets. Cell Rep. 2017;21: 2597–2613.

63. Kirkpatrick RM, McGue M, Iacono WG, Miller MB, Basu S. Results of a "GWAS plus:" general cognitive ability is substantially heritable and massively polygenic. PLoS One. 2014;9: e112390.

64. Rietveld CA, Medland SE, Derringer J, Yang J, Esko T, Martin NW, et al. GWAS of 126,559 individuals identifies genetic variants associated with educational attainment. Science. 2013;340: 1467–1471.

65. Lee JJ, Wedow R, Okbay A, Kong E, Maghzian O, Zacher M, et al. Gene discovery and polygenic prediction from a genome-wide association study of educational attainment in 1.1 million individuals. Nat Genet. 2018;50: 1112–1121.

66. Greenspan NS. Genes, Heritability, "Race", and Intelligence: Misapprehensions and Implications. Genes . 2022;13. doi:10.3390/genes13020346

67. Turkheimer E, Harden KP, Nisbett RE. There's still no good reason to believe black-white IQ differences are due to genes. In: Vox [Internet]. 15 Jun 2017 [cited 6 Mar 2023]. Available: https://www.vox.com/the-big-idea/2017/6/15/15797120/race-black-white-iq-response-critics

68. 1000 Genomes Project Consortium, Auton A, Brooks LD, Durbin RM, Garrison EP, Kang HM, et al. A global reference for human genetic variation. Nature. 2015;526: 68–74.

69. Roth WD, Ivemark B. Genetic Options: The Impact of Genetic Ancestry Testing on Consumers' Racial and Ethnic Identities1. Am J Sociol. 2018 [cited 18 Jan 2023]. doi:10.1086/697487

70. Davis LK. Human Genetics Needs an Antiracism Plan. In: Scientific American [Internet]. [cited 28 Feb 2023]. Available: https://www.scientificamerican.com/article/human-genetics-needs-an-antiracism-plan/

71. Ashley Smart U. A Field at a Crossroads: Genetics and Racial Mythmaking. Scientific American. By Ashley Smart, Undark on December 22 2022. Available: https://www.scientificamerican.com/article/a-field-at-a-crossroads-genetics-and-racial-mythmaking/. Accessed 18 Jan 2023.

72. Popejoy AB. Too many scientists still say Caucasian. In: Nature Publishing Group UK [Internet]. 24 Aug 2021 [cited 25 Jun 2022]. doi:10.1038/d41586-021-02288-x

73. Byeon YJJ, Islamaj R, Yeganova L, Wilbur WJ, Lu Z, Brody LC, et al. Evolving use of ancestry, ethnicity, and race in genetics research-A survey spanning seven decades. Am J Hum Genet. 2021;108: 2215–2223.

74. Wills M. Are Clusters Races? A Discussion of the Rhetorical Appropriation of Rosenberg et al.'s "Genetic Structure of Human Populations." Philosophy & Theory in Biology. 2017;9. doi:10.3998/ptb.6959004.0009.012

75. Anderson W. ASHG Documents and Apologizes for Past Harms of Human Genetics Research, Commits to Building an Equitable Future. In: ASHG [Internet]. 24 Jan 2023 [cited 26 Feb 2023]. Available: https://www.ashg.org/publications-news/press-releases/ashg-documents-and-apologizes-for-past-harms-of-human-genetics-research-commits-to-building-an-equitable-future/

76. American Academy of Pediatrics Board of Directors and Executive Committee. AAP Perspective: Race-Based Medicine. Pediatrics. 2021;148. doi:10.1542/peds.2021-053829

77. Amutah C, Greenidge K, Mante A, Munyikwa M, Surya SL, Higginbotham E, et al. Misrepresenting Race — The Role of Medical Schools in Propagating Physician Bias. New England Journal of Medicine. 2021. pp. 872–878. doi:10.1056/nejmms2025768

78. Tishkoff SA, Verrelli BC. Patterns of human genetic diversity: implications for human evolutionary history and disease. Annu Rev Genomics Hum Genet. 2003;4: 293–340.

79. Waples RS, Gaggiotti O. What is a population? An empirical evaluation of some genetic methods for identifying the number of gene pools and their degree of connectivity. Mol Ecol. 2006;15: 1419–1439.

80. Timpson NJ, Greenwood CMT, Soranzo N, Lawson DJ, Richards JB. Genetic architecture: the shape of the genetic contribution to human traits and disease. Nat Rev Genet. 2017;19: 110–124.

81. Variants Associated with Common Disease Are Not Unusually Differentiated in Frequency across Populations. Am J Hum Genet. 2006;78: 130–136.

82. Marigorta UM, Navarro A. High Trans-ethnic Replicability of GWAS Results Implies Common Causal Variants. PLoS Genet. 2013;9: e1003566.

83. Li YR, Keating BJ. Trans-ethnic genome-wide association studies: advantages and challenges of mapping in diverse populations. Genome Med. 2014;6: 1–14.

84. Wojcik GL, Graff M, Nishimura KK, Tao R, Haessler J, Gignoux CR, et al. Genetic analyses of diverse populations improves discovery for complex traits. Nature. 2019;570: 514–518.

85. Global Biobank Meta-analysis Initiative (GBMI) consortium. [cited 6 Mar 2023]. Available: https://www.cell.com/consortium/global-biobank-meta-analysis-initiative

86. Transethnic Genetic-Correlation Estimates from Summary Statistics. Am J Hum Genet. 2016;99: 76–88.

87. Momin MM, Shin J, Lee S, Truong B, Benyamin B, Lee SH. A method for an unbiased estimate of cross-ancestry genetic correlation using individual-level data. Nat Commun. 2023;14: 722.

88. Guo J, Bakshi A, Wang Y, Jiang L, Yengo L, Goddard ME, et al. Quantifying genetic heterogeneity between continental populations for human height and body mass index. Sci Rep. 2021;11: 5240.

89. Galinsky KJ, Reshef YA, Finucane HK, Loh P-R, Zaitlen N, Patterson NJ, et al. Estimating cross-population genetic correlations of causal effect sizes. Genet Epidemiol. 2019;43: 180–188.

90. Lam M, Chen C-Y, Li Z, Martin AR, Bryois J, Ma X, et al. Comparative genetic architectures of schizophrenia in East Asian and European populations. Nat Genet. 2019;51: 1670–1678.

91. Shi H, Gazal S, Kanai M, Koch EM, Schoech AP, Siewert KM, et al. Population-specific causal disease effect sizes in functionally important regions impacted by selection. Nat Commun. 2021;12: 1–15.

92. Hou K, Ding Y, Xu Z, Wu Y, Bhattacharya A, Mester R, et al. Causal effects on complex traits are similar for common variants across segments of different continental ancestries within admixed individuals. Nat Genet. 2023. doi:10.1038/s41588-023-01338-6

93. Dudley JT, Kim Y, Liu L, Markov GJ, Gerold K, Chen R, et al. Human genomic disease variants: a neutral evolutionary explanation. Genome Res. 2012;22: 1383–1394.

94. Nei M, Suzuki Y, Nozawa M. The neutral theory of molecular evolution in the genomic era. Annu Rev Genomics Hum Genet. 2010;11: 265–289.

95. Campbell MC, Tishkoff SA. African genetic diversity: implications for human demographic history, modern human origins, and complex disease mapping. Annu Rev Genomics Hum Genet. 2008;9: 403–433.

96. Kim MS, Patel KP, Teng AK, Berens AJ, Lachance J. Genetic disease risks can be misestimated across global populations. Genome Biol. 2018;19: 179.

97. Gurdasani D, Carstensen T, Fatumo S, Chen G, Franklin CS, Prado-Martinez J, et al. Uganda Genome Resource Enables Insights into Population History and Genomic Discovery in Africa. Cell. 2019;179: 984–1002.e36.

98. Zeng J, Xue A, Jiang L, Lloyd-Jones LR, Wu Y, Wang H, et al. Widespread signatures of natural selection across human complex traits and functional genomic categories. Nat Commun. 2021;12: 1164.

99. Zeng J, de Vlaming R, Wu Y, Robinson MR, Lloyd-Jones LR, Yengo L, et al. Signatures of negative selection in the genetic architecture of human complex traits. Nat Genet. 2018;50: 746–753.

100. O'Connor LJ, Schoech AP, Hormozdiari F, Gazal S, Patterson N, Price AL. Extreme Polygenicity of Complex Traits Is Explained by Negative Selection. Am J Hum Genet. 2019;105: 456–476.

101. Schoech AP, Jordan DM, Loh P-R, Gazal S, O'Connor LJ, Balick DJ, et al. Quantification of frequency-dependent genetic architectures in 25 UK Biobank traits reveals action of negative selection. Nat Commun. 2019;10: 790.

102. Pardiñas AF, Holmans P, Pocklington AJ, Escott-Price V, Ripke S, Carrera N, et al. Common schizophrenia alleles are enriched in mutation-intolerant genes and in regions under strong background selection. Nat Genet. 2018;50: 381–389.

103. Evidence for Variable Selective Pressures at MC1R. Am J Hum Genet. 2000;66: 1351–1361.

104. Rana BK, Hewett-Emmett D, Jin L, Chang BH, Sambuughin N, Lin M, et al. High polymorphism at the human melanocortin 1 receptor locus. Genetics. 1999;151: 1547–1557.

105. Worldwide polymorphism at the MC1R locus and normal pigmentation variation in humans. Peptides. 2005;26: 1901–1908.

106. John PR, Makova K, Li W-H, Jenkins T, Ramsay M. DNA polymorphism and selection at the melanocortin-1 receptor gene in normally pigmented southern African individuals. Ann N Y Acad Sci. 2003;994: 299–306.

107. Slatkin M. Linkage disequilibrium — understanding the evolutionary past and mapping the medical future. Nature Reviews Genetics. 2008. pp. 477–485. doi:10.1038/nrg2361

108. Barton NH, Otto SP. Evolution of recombination due to random drift. Genetics. 2005;169: 2353–2370.

109. Duan Q, Liu EY, Auer PL, Zhang G, Lange EM, Jun G, et al. Imputation of coding variants in African Americans: better performance using data from the exome sequencing project. Bioinformatics. 2013;29: 2744–2749.

110. Liu EY, Li M, Wang W, Li Y. MaCH-admix: genotype imputation for admixed populations. Genet Epidemiol. 2013;37: 25–37.

111. Vergara C, Parker MM, Franco L, Cho MH, Valencia-Duarte AV, Beaty TH, et al. Genotype imputation performance of three reference panels using African ancestry individuals. Hum Genet. 2018;137: 281–292.

112. Schaid DJ, Chen W, Larson NB. From genome-wide associations to candidate causal variants by statistical fine-mapping. Nat Rev Genet. 2018;19: 491–504.

113. Kraft P, Zeggini E, Ioannidis JPA. Replication in genome-wide association studies. Stat Sci. 2009;24: 561.

114. LaPierre N, Taraszka K, Huang H, He R, Hormozdiari F, Eskin E. Identifying causal variants by fine mapping across multiple studies. PLoS Genet. 2021;17: e1009733.

115. Kichaev G, Pasaniuc B. Leveraging Functional-Annotation Data in Trans-ethnic Fine-Mapping Studies. Am J Hum Genet. 2015;97: 260.

116. Jolliffe IT. Principal Component Analysis and Factor Analysis. In: Jolliffe IT, editor. Principal Component Analysis. New York, NY: Springer New York; 1986. pp. 115–128.

117. McVean G. A genealogical interpretation of principal components analysis. PLoS Genet. 2009;5: e1000686.

118. Novembre J, Johnson T, Bryc K, Kutalik Z, Boyko AR, Auton A, et al. Genes mirror geography within Europe. Nature. 2008;456: 98–101.

119. Dauda B, Molina SJ, Allen DS, Fuentes A, Ghosh N, Mauro M, et al. Ancestry: How researchers use it and what they mean by it. Front Genet. 2023;14. doi:10.3389/fgene.2023.1044555

120. Coop G. Genetic similarity versus genetic ancestry groups as sample descriptors in human genetics. arXiv [q-bio.PE]. 2022. Available: http://arxiv.org/abs/2207.11595

121. Need AC, Kasperavičiūtė D, Cirulli ET, Goldstein DB. A genome-wide genetic signature of Jewish ancestry perfectly separates individuals with and without full Jewish ancestry in a large random sample of European Americans. Genome Biol. 2009;10: 1–7.

122. Falk R. Genetic markers cannot determine Jewish descent. Front Genet. 2015;5. doi:10.3389/fgene.2014.00462

123. Tofanelli S, Taglioli L, Bertoncini S, Francalacci P, Klyosov A, Pagani L. Mitochondrial and Y chromosome haplotype motifs as diagnostic markers of Jewish ancestry: a reconsideration. Front Genet. 2014;5. doi:10.3389/fgene.2014.00384

124. Elhaik E. In Search of the jüdische Typus: A Proposed Benchmark to Test the Genetic Basis of Jewishness Challenges Notions of "Jewish Biomarkers." Front Genet. 2016;7. doi:10.3389/fgene.2016.00141

125. Carlson J, Harris K. Quantifying and contextualizing the impact of bioRxiv preprints through automated social media audience segmentation. PLoS Biol. 2020;18: e3000860.

126. Website. doi:10.1073/pnas.1600398113

127. López S, Tarekegn A, Band G, van Dorp L, Bird N, Morris S, et al. Evidence of the interplay of genetics and culture in Ethiopia. Nat Commun. 2021;12: 1–15.

128. Baker JL, Rotimi CN, Shriner D. Human ancestry correlates with language and reveals that race is not an objective genomic classifier. Sci Rep. 2017;7: 1–10.

129. Jablonski NG. Human skin pigmentation as an example of adaptive evolution. Proc Am Philos Soc. 2012;156: 45–57.

130. Jablonski NG, Chaplin G. Human skin pigmentation as an adaptation to UV radiation. Proceedings of the National Academy of Sciences. 2010;107: 8962–8968.

131. Deng L, Xu S. Adaptation of human skin color in various populations. Hereditas. 2018;155: 1.

132. Mckusick VA, Eldridge R, Hostetler JA, Egeland JA. DWARFISM IN THE AMISH. Trans Assoc Am Physicians. 1964;77: 151–168.

133. Strauss KA, Puffenberger EG. Genetics, medicine, and the Plain people. Annu Rev Genomics Hum Genet. 2009;10: 513–536.

134. Plain Community Translational Medicine Program and Clinic. In: Children's Hospital of Pittsburgh [Internet]. [cited 12 Apr 2023]. Available: https://www.chp.edu/our-services/genetics/clinical-services/plain-communities-clinic

135. Clinic for Special Children. In: Clinic for Special Children [Internet]. [cited 12 Apr 2023]. Available: https://clinicforspecialchildren.org/

136. Medina-Gomez C, Felix JF, Estrada K, Peters MJ, Herrera L, Kruithof CJ, et al. Challenges in conducting genome-wide association studies in highly admixed multi-ethnic populations: the Generation R Study. Eur J Epidemiol. 2015;30: 317–330.

137. Website. doi:10.1073/pnas.0126614100

138. Magalhães da Silva T, Sandhya Rani MR, de Oliveira Costa GN, Figueiredo MA, Melo PS, Nascimento JF, et al. The correlation between ancestry and color in two cities of Northeast Brazil with contrasting ethnic compositions. Eur J Hum Genet. 2014;23: 984–989.

139. Leite TKM, Fonseca RMC, de França NM, Parra EJ, Pereira RW. Genomic Ancestry, Self-Reported "Color" and Quantitative Measures of Skin Pigmentation in Brazilian Admixed Siblings. PLoS One. 2011;6: e27162.

140. Klinger EV, Carlini SV, Gonzalez I, Hubert SS, Linder JA, Rigotti NA, et al. Accuracy of Race, Ethnicity, and Language Preference in an Electronic Health Record. J Gen Intern Med. 2014;30: 719–723.

141. Taliun D, Harris DN, Kessler MD, Carlson J, Szpiech ZA, Torres R, et al. Sequencing of 53,831 diverse genomes from the NHLBI TOPMed Program. Nature. 2021;590: 290–299.

142. Kowalski MH, Qian H, Hou Z, Rosen JD, Tapia AL, Shan Y, et al. Use of >100,000

NHLBI Trans-Omics for Precision Medicine (TOPMed) Consortium whole genome sequences improves imputation quality and detection of rare variant associations in admixed African and Hispanic/Latino populations. PLoS Genet. 2019;15: e1008500.

143. Peterson RE, Kuchenbaecker K, Walters RK, Chen C-Y, Popejoy AB, Periyasamy S, et al. Genome-wide Association Studies in Ancestrally Diverse Populations: Opportunities, Methods, Pitfalls, and Recommendations. Cell. 2019;179: 589–603.

144. Aspinall PJ. Ethnic/Racial Terminology as a Form of Representation: A Critical Review of the Lexicon of Collective and Specific Terms in Use in Britain. Genealogy. 2020;4: 87.

145. Website. Available: https://www.journals.uchicago.edu/doi/full/10.1086/644532

146. Zhou X, Stephens M. Genome-wide efficient mixed-model analysis for association studies. Nat Genet. 2012;44: 821–824.

147. Kang HM, Sul JH, Service SK, Zaitlen NA, Kong S-Y, Freimer NB, et al. Variance component model to account for sample structure in genome-wide association studies. Nat Genet. 2010;42: 348–354.

148. Listgarten J, Lippert C, Kadie CM, Davidson RI, Eskin E, Heckerman D. Improved linear mixed models for genome-wide association studies. Nat Methods. 2012;9: 525–526.

149. Loh P-R, Tucker G, Bulik-Sullivan BK, Vilhjálmsson BJ, Finucane HK, Salem RM, et al. Efficient Bayesian mixed-model analysis increases association power in large cohorts. Nat Genet. 2015;47: 284–290.

150. Loh P-R, Kichaev G, Gazal S, Schoech AP, Price AL. Mixed-model association for biobank-scale datasets. Nat Genet. 2018;50: 906–908.

151. Zhou W, Nielsen JB, Fritsche LG, Dey R, Gabrielsen ME, Wolford BN, et al. Efficiently controlling for case-control imbalance and sample relatedness in large-scale genetic association studies. Nat Genet. 2018;50: 1335–1341.

152. Lin D-Y, Tao R, Kalsbeek WD, Zeng D, Gonzalez F, Fernández-Rhodes L, et al. Genetic Association Analysis under Complex Survey Sampling: The Hispanic Community Health Study/Study of Latinos. Am J Hum Genet. 2014;95: 675–688.

153. Hu Y, Bien SA, Nishimura KK, Haessler J, Hodonsky CJ, Baldassari AR, et al. Multi-ethnic genome-wide association analyses of white blood cell and platelet traits in the Population Architecture using Genomics and Epidemiology (PAGE) study. BMC Genomics. 2021;22. doi:10.1186/s12864-021-07745-5

154. Conomos MP, Reiner AP, McPeek MS, Thornton TA. Genome-Wide Control of Population Structure and Relatedness in Genetic Association Studies via Linear Mixed Models with Orthogonally Partitioned Structure. bioRxiv. 2018. p. 409953. doi:10.1101/409953

155. [No title]. In: Liebertpub.com [Internet]. [cited 3 Feb 2023]. Available: https://www.liebertpub.com/doi/abs/

156. Sul JH, Martin LS, Eskin E. Population structure in genetic studies: Confounding factors and mixed models. PLoS Genet. 2018;14: e1007309.

157. Website. Available: https://onlinelibrary.wiley.com/doi/10.1002/gepi.22384

158. Wolford BN, Willer CJ, Surakka I. Electronic health records: the next wave of complex disease genetics. Hum Mol Genet. 2018;27: R14–R21.

159. Edwards BJ, Haynes C, Levenstien MA, Finch SJ, Gordon D. Power and sample size calculations in the presence of phenotype errors for case/control genetic association studies. BMC Genet. 2005;6: 18.

160. Barnes A. Race, schizophrenia, and admission to state psychiatric hospitals. Adm Policy Ment Health. 2004;31: 241–252.

161. Bigdeli TB, Voloudakis G, Barr PB, Gorman BR, Genovese G, Peterson RE, et al. Penetrance and Pleiotropy of Polygenic Risk Scores for Schizophrenia, Bipolar Disorder, and Depression Among Adults in the US Veterans Affairs Health Care System. JAMA Psychiatry. 2022;79: 1092–1101.

162. Schwartz RC, Blankenship DM. Racial disparities in psychotic disorder diagnosis: A review of empirical literature. World Journal of Psychiatry. 2014;4: 133.

163. Barnes A. Race and hospital diagnoses of schizophrenia and mood disorders. Soc Work. 2008;53: 77–83.

164. Kirby JC, Speltz P, Rasmussen LV, Basford M, Gottesman O, Peissig PL, et al. PheKB: a catalog and workflow for creating electronic phenotype algorithms for transportability. J Am Med Inform Assoc. 2016;23: 1046–1052.

165. Himmelstein G, Bates D, Zhou L. Examination of Stigmatizing Language in the Electronic Health Record. JAMA Netw Open. 2022;5: e2144967.

166. [No title]. [cited 6 Mar 2023]. Available: https://arxiv.org/pdf/2108.01764.pdf

167. Hoffman KM, Trawalter S, Axt JR, Oliver MN. Racial bias in pain assessment and treatment recommendations, and false beliefs about biological differences between blacks and whites. Proc Natl Acad Sci U S A. 2016;113: 4296–4301.

168. Bear Don't Walk OJ 4th, Reyes Nieva H, Lee SS-J, Elhadad N. A scoping review of ethics considerations in clinical natural language processing. JAMIA Open. 2022;5: ooac039.

169. Sikstrom L, Maslej MM, Hui K, Findlay Z, Buchman DZ, Hill SL. Conceptualizing fairness: three pillars for medical algorithms and health equity. BMJ Health Care Inform.

2022;29. doi:10.1136/bmjhci-2021-100459

170. Thompson HM, Sharma B, Bhalla S, Boley R, McCluskey C, Dligach D, et al. Bias and fairness assessment of a natural language processing opioid misuse classifier: detection and mitigation of electronic health record data disadvantages across racial subgroups. J Am Med Inform Assoc. 2021;28: 2393–2403.

171. Dueñas HR, Seah C, Johnson JS, Huckins LM. Implicit bias of encoded variables: frameworks for addressing structured bias in EHR–GWAS data. Hum Mol Genet. 2020;29: R33–R41.

172. Thorlby R, Jorgensen S, Siegel B, Ayanian JZ. How Health Care Organizations Are Using Data on Patients' Race and Ethnicity to Improve Quality of Care. Milbank Q. 2011;89: 226.

173. Liz J. "The Fixity of Whiteness": Genetic Admixture and the Legacy of the One-Drop Rule. Critical Philosophy of Race. 2018;6: 239–261.

174. Lujan HL, DiCarlo SE. The racist "one drop rule" influencing science: it is time to stop teaching "race corrections" in medicine. Adv Physiol Educ. 2021;45. doi:10.1152/advan.00063.2021

175. Hickman CB. The Devil and the One Drop Rule: Racial Categories, African Americans, and the U.S. Census. Mich Law Rev. 1997;95: 1161–1265.

176. Bower JK, Patel S, Rudy JE, Felix AS. Addressing Bias in Electronic Health Record-Based Surveillance of Cardiovascular Disease Risk: Finding the Signal Through the Noise. Curr Epidemiol Rep. 2017;4: 346–352.

177. Website. Available: https://www.jstor.org/stable/26556356

178. Goldstein E, Topitzes J, Miller-Cribbs J, Brown RL. Influence of race/ethnicity and income on the link between adverse childhood experiences and child flourishing. Pediatr Res. 2020;89: 1861–1869.

179. Parolin Z, Lee EK. The Role of Poverty and Racial Discrimination in Exacerbating the Health Consequences of COVID-19. Lancet Reg Health Am. 2022;7: 100178.

180. Gage SH, Smith GD, Ware JJ, Flint J, Munafò MR. G = E: What GWAS Can Tell Us about the Environment. PLoS Genet. 2016;12: e1005765.

181. Chen M, Tan X, Padman R. Social determinants of health in electronic health records and their impact on analysis and risk prediction: A systematic review. J Am Med Inform Assoc. 2020;27: 1764–1773.

182. Website. doi:10.1146/annurev-publhealth-032315-021353

183. Roth C, Foraker RE, Payne PRO, Embi PJ. Community-level determinants of obesity:

harnessing the power of electronic health records for retrospective data analysis. BMC Med Inform Decis Mak. 2014;14: 1–8.

184. Boland MR, Davidson LM, Canelón SP, Meeker J, Penning T, Holmes JH, et al. Harnessing electronic health records to study emerging environmental disasters: a proof of concept with perfluoroalkyl substances (PFAS). NPJ Digit Med. 2021;4: 122.

185. Committee on the Recommended Social and Behavioral Domains and Measures for Electronic Health Records, Board on Population Health and Public Health Practice, Institute of Medicine. Summary. Capturing Social and Behavioral Domains in Electronic Health Records: Phase 1. National Academies Press (US); 2014.

186. Committee on the Recommended Social and Behavioral Domains and Measures for Electronic Health Records, Board on Population Health and Public Health Practice, Institute of Medicine. Recommended Core Domains and Measures. Capturing Social and Behavioral Domains and Measures in Electronic Health Records: Phase 2. National Academies Press (US); 2015.

187. Kasthurirathne SN, Vest JR, Menachemi N, Halverson PK, Grannis SJ. Assessing the capacity of social determinants of health data to augment predictive models identifying patients in need of wraparound social services. J Am Med Inform Assoc. 2018;25: 47–53.

188. Vest JR, Ben-Assuli O. Prediction of emergency department revisits using area-level social determinants of health measures and health information exchange information. Int J Med Inform. 2019;129: 205–210.

189. Bhavsar NA, Gao A, Phelan M, Pagidipati NJ, Goldstein BA. Value of Neighborhood Socioeconomic Status in Predicting Risk of Outcomes in Studies That Use Electronic Health Record Data. JAMA Netw Open. 2018;1: e182716.

| Biobank | Healthcare system | White | Black/ African American | Hispanic/ Latino | Asian | Other | Ref. |
|---|---|---|---|---|---|---|---|
| BioMe Biobank | Mount Sinai Health System | 27.06 | 22.29 | 2.49 | 33.95 | 14.21 | [25] |
| BioVu | Vanderbilt University Medical Center | 85.00 | 11.00 | 1.00 | 1.00 | 2.00 | [26] |
| Colorado Center for Personalized Medicine Biobank | UCHealth | 83.03 | 3.80 | 2.08 | 9.22 | 1.87 | [27] |
| Geisinger MyCode Community Health Initiative | Geisinger Health System | 92.60 | 3.70 | 3.20 | Not reported | 0.50 | [28], [29] |
| Mayo Clinic Biobank | Mayo Clinic Health System | 95.00 | 0.50 | Not reported | 1.20 | 3.30 | [30] |
| Mass General Brigham Biobank | Mass General Brigham | 82.10 | 5.20 | 2.60 | 5.90 | 4.20 | [31] |
| Michigan Genomics Initiative | Michigan Medicine | 86.94 | 5.65 | 2.32 | 1.41 | 3.68 | [32] |
| Genetic Epidemiology Research on Adult Health and Aging Cohort | Kaiser Permanente Northern California | 78.10 | 3.20 | 7.10 | 7.10 | 4.50 | [33] |
| UCLA ATLAS Community Health Initiative | UCLA Health | 55.18 | 6.59 | 10.31 | 17.20 | 10.72 | [34] |
| Penn Medicine Biobank | Penn Medicine | 71.20 | 16.80 | 4.10 | 3.30 | 4.60 | [35] |

**Supplementary Table 1:** Summary of EHR-linked biobanks embedded within healthcare systems within the United States. Columns show the self-identified race and ethnicity percentages per biobank. Because the Geisinger cohort description reference did not include statistics by race and ethnicity, we report the statistics for their active health system patients as reported in [29].